\documentclass[12pt]{article}
\usepackage{graphicx}
\usepackage{subfigure}
\input epsf.tex

\newcommand{\beq}{\begin{equation}}
\newcommand{\eeq}{\end{equation}}
\newcommand{\be}{\begin{equation}}
\newcommand{\ee}{\end{equation}}
\newcommand{\bea}{\begin{eqnarray}}
\newcommand{\eea}{\end{eqnarray}}

\newcommand{\gs}{\mbox{$g_s$}}            
\newcommand{\ap}{\mbox{$\alpha^\prime$}}  
\newcommand{\ls}{\mbox{$l_s$}}            

\def\href#1#2{#2}

\def\p{\partial}

\begin{document}

\baselineskip=15.5pt
\pagestyle{plain}
\setcounter{page}{1}

\begin{titlepage}
\begin{flushleft}
       \hfill                      {\tt hep-th/0907.1170}\\
       \hfill                       FIT HE - 09-01\\
       \hfill                       Kagoshima HE - 09-1 \\
\end{flushleft}
\vspace*{3mm}

\begin{center}
  {\huge Stability of D brane Anti D brane Systems\\
  \vspace*{2mm}
in Confining Gauge Theories}
\end{center}
\vspace{5mm}

\begin{center}
\vspace*{5mm}
\vspace*{2mm}
\vspace*{5mm}
{\large Kazuo Ghoroku${}^{\dagger}$\footnote[1]{\tt gouroku@dontaku.fit.ac.jp},
\vspace*{5mm}
${}^{\S}$Akihiro Nakamura\footnote[2]{\tt nakamura@sci.kagoshima-u.ac.jp}
${}^{\P}$Fumihiko Toyoda\footnote[3]{\tt ftoyoda@fuk.kindai.ac.jp}
%
}\\

{${}^{\dagger}$Fukuoka Institute of Technology, Wajiro, 
Higashi-ku} \\
{
Fukuoka 811-0295, Japan\\}
{
${}^{\S}$Department of Physics, Kagoshima University, Korimoto 1-21-35, \\Kagoshima 890-0065, Japan\\}
{
${}^{\P}$School of Humanity-Oriented Science and
Engineering, Kinki University,\\ Iizuka 820-8555, Japan}

\vspace*{10mm}
\end{center}

\begin{center}
{\large Abstract}
\end{center}
We study the stability of a special form of D brane embedding which is regarded as
a bound state of D$_n$ and anti-D$_n$-brane embedded in a 10D supergravity
background which is dual to a confining gauge theory. 
For D5 branes with $U(1)$ flux, their bound state configuration can be regarded
as the baryonium vertex. For D branes of $n=6$ and 8 without the $U(1)$ flux,
their bound states have been used to introduce flavor quarks in the dual supersymmetric
Yang-Mills theory.
In any case, it would be important to assure that they are free from tachyon
instability. For all these cases, we could show their stability with respect to this point. 

\noindent

\vfill
\begin{flushleft}

\end{flushleft}
\end{titlepage}
\newpage

\section{Introduction}

It would be a challenging problem to make clear the deep structure of
mesons, baryons and other hadrons which are the bound states of quarks.
The quantum chromodynamics is well defined to solve these problems, but
it is very difficult to see the non-perturbative properties of this theory.
A very promising approach in this direction has been performed in the recent holographic
approaches. According to the quark model, the baryons 
are composed of $N_c$ quarks combined at a vertex point, which would however
not a point but has a structure
and the vertex with a structure has been
expressed in type IIB theory by 
D5-branes wrapped on $S^5$ in AdS${}_5\times S^5$ space-time 
\cite{wittenbaryon,groguri,imamura,cgs,GRST,cgst,ima-04,baryonsugra,imafirst,CGMP,MN,JLG} 
based on the string/gauge theory correspondence
\cite{jthroat,gkpads,wittenholo}. 
And the baryon has been constructed as a system of fundamental 
strings (F-strings) corresponding to the quarks and this vertex
\cite{GI,GINT,ALR,BLL,LZP,SekiS,BCH}.

In such models, the F-strings are dissolved as a $U(1)$ flux in the D5 brane, and 
their remaining parts flow out from $S^5$ through one (or two) cusp(s) 
on the surface of the D5 brane to the outside of the extended vertex. The outside
fluxes are
separated as $N_c$ free strings which are connected to the same number of
separated quarks. The vertex has been 
embedded on $S^5$ in 10D bulk, and its shape could be deformed
due to the dissolved $U(1)$ flux and also by the 10D bulk backbacground which is dual to the YM gauge theory in the confinement phase. This deformation can also be seen in our
real 3D space as a string like object \cite{imamura,cgs,GRST,cgst}.

As a special deformed solution of the embedded D5 brane, we 
have found a configuration which corresponds to a bound state of D5 and anti-D5 brane
\cite{GINT2}.
And this configuration has been regarded as a baryonium (a bound state of baryon and 
anti-baryon) vertex 
since this has two cusps where a definite number ($N_q$) of $U(1)$ fluxes enter from
one side and the same number of fluxes come out from the other cusp. So this vertex
connects $N_q$ quarks and the same number of anti-quarks. So this state can be considered
as a meson with a special vertex, and it is 
called as baryonium. So our holographic approach have a chance to resolve the long standing problem whether this kind of state could be realized or forbidden. So it is an important
issue to study the stability of our baryonium configuration. 

Previously, we have resolved this problem from the view point of the energy, and
we find that there is a minimum point of the classical action for an appropriate 
solutions of the baryonium. However,
in general, we should be
aware of an instability coming from tachyons which are 
anticipated in this kind of configurations, the bound state composed of D and anti-D branes.
This time, we examine this point by studying the spectrum of the fluctuation
for the field living on the D5 brane to search for the
tachyonic modes. 

We also study the stability of the other cases of the D and anti-D brane bound state
used in the other holographic gauge theory.

\vspace{.3cm}
In the next section, we give our model and D5-brane action
with non-trivial $U(1)$ gauge field, and
the baryonium solutions are briefly reviewed.
In section 3, we give a method to study the eigenvalue of the fluctuation modes
and the stability of the baryonium is discussed. The 
stability of the bound states of D8 and anti-D8,
and D6 and anti-D6 branes, which are considered to introduce fundamental quarks, are
also studied by the same method in section 4.
Finally in the final section, we summarize our results and discuss related
problems.

\section{$D5/\overline{D5}$ and Baryonium}\label{eqnsec}
The $D5$-$\overline{D5}$ configurations have been proposed
as baryonium vertex. The configurations have been given
as solutions of the equations of motion of the D5-brane which 
is embedded as a probe in a supersymmetric 10d background of type IIB theory. 
The dual theory of this background corresponds to the confining gauge theory,
then the configuration given here can be considered as a bound state of baryon
and anti-baryon. How to obtain this solution is briefly reviewed in the following.

\subsection{Bulk background and $D5$ brane action}
 
We consider the following supergravity background solution
\cite{Liu:1999fc,KS2,GY},
\beq
ds^2_{10}= e^{\Phi/2}
\left(
\frac{r^2}{R^2}\eta_{\mu\nu}dx^\mu dx^\nu +
\frac{R^2}{r^2} dr^2+R^2 d\Omega_5^2 \right) \ ,
\label{non-SUSY-sol}
\eeq
which is written in string frame. And the dilaton $\Phi$ and
the axion $\chi$ are given as
\beq 
 e^\Phi= 
1+\frac{q}{r^4} \ , \quad \chi=-e^{-\Phi}+\chi_0 \ ,
\label{dilaton} 
\eeq
with self-dual Ramond-Ramond field strength
\begin{equation}\label{fiveform}
C_{(5)}\equiv dC_{(4)}=4R^4\left(\mbox{vol}(S^5) d\theta_1\wedge\ldots\wedge d\theta_5
-{r^3\over R^8}  dt\wedge \ldots\wedge dx_3\wedge dr\right),
\end{equation}
where $\mbox{vol}(S^5)\equiv\sin^4\theta_1
\mbox{vol}(S^4)\equiv\sin^4\theta_1\sin^3\theta_2\sin^2\theta_3\sin\theta_4$.

This solution, (\ref{non-SUSY-sol})-(\ref{fiveform}),
is useful since the confinement
of quarks are realized due to the gauge condensate $
q\equiv \langle F_{\mu\nu}^2\rangle$ \cite{KS2,GY}, which is given by the coefficient
of $1/r^4$ for the asymptotic expansion of the dilaton at large $r$.
And furthermore, $\cal{N}$=2 supersymmetry is preserved in spite of the non-trivial
dilaton is introduced.
We can assure through the Wilson loop that $q^{1/2}$ is proportional to
the tention of the linear rising potential between the quark 
and anti-quark \cite{GY}. In the present case, $q$ is essential to fix
the size of the baryonium and stabilize it energetically. 
We notice that
the axion $\chi$ corresponds to the source of D(-1) brane and it is Wick rotated
in the supergravity action. This is necessary to preserve the supersymmetry.

\label{eqnsec2}

In the next, we introduce the probe D5 brane, and its action must include
$N_c$ dissolved $U(1)$ fluxes in it.
The D5-brane action 
is thus written as by the Dirac-Born-Infeld (DBI) plus
WZW term \cite{cgs}
\begin{eqnarray}\label{d5action}
S_{D5}&=&-T_{5}\int d^6\xi
 e^{-\Phi}\sqrt{-\det\left(g_{ab}+\tilde{F}_{ab}\right)}+T_{5}\int
d^6\xi \tilde{A}_{(1)}\wedge \tilde{C}_{(5)} ~,\\
g_{ab}&\equiv&\p_a X^{\mu}\p_b X^{\nu}G_{\mu\nu}~, \qquad
\tilde{C}_{a_1\ldots
a_5}\,\equiv\,\p_{a_1}X^{\mu_1}\ldots\p_{a_5}X^{\mu_5}C_{\mu_1\ldots\mu_5}~.\nonumber
\end{eqnarray}
where $T_5=1/(\gs(2\pi)^{5}\ls^{6})$ is the brane tension,
and $U(1)$ worldvolume field strength is expressed by
$\tilde{F}_{ab}=d\tilde{A}_{(1)}$ with $\tilde{F}_{ab}/2\pi\ap =F_{ab}
=d{A}_{(1)}$. And $\tilde{C}_{(5)}$ denotes the induced five form field strength.
.

\subsection{Equations of motion and Baryonium solution}\label{bound-2}

The D5 brane is embedded in the world volume
$\xi^{a}=(t,\theta,\theta_2,\ldots,\theta_5)$, where $(\theta_2,\ldots,\theta_5)$
are the $S^4$ part with the volume of $\Omega_{4}=8\pi^{2}/3$, where we set
as $\theta_1=\theta$. 
Then, we restrict our attention to $SO(5)$ symmetric configurations of the 
form $r(\theta)$, $x(\theta)$, and $A_t(\theta)$ (with all other fields 
set to zero). 
In this case, the above action is written as  \cite{cgst}
\be \label{d3action}
S= T_5 \Omega_{4}R^4\int dt\,d\theta \{ -\sin^4\theta 
  \sqrt{e^{\Phi}\left(r^2+r^{\prime 2}+(r/R)^{4}x^{\prime 2}\right)
   -\tilde{F}_{t \theta}^2}  -\tilde{F}_{t \theta} D \},
\ee
where the WZW term is rewritten by partial integration with respect to
$\theta$, and $\Omega_{4}=8\pi^{2}/3$ is the volume of the unit four-sphere.
The factor $D(\theta)$ is defined by
\beq\label{D}
\partial_\theta D = -4 \sin^4\theta~,
\eeq
and is related to $\tilde{F}_{t \theta}$ by the equation of motion for
$\tilde{A}_{t}$ given below. 
We call this $D$ as displacement. It is given by 
solving (\ref{D}) as follows ,
\be \label{d}
  D(\nu,\theta) \equiv \left[{3\over 2}(\nu\pi-\theta)
  +{3\over 2}\sin\theta\cos\theta+\sin^{3}\theta\cos\theta\right].
\ee
Here $\nu$ is an integration constant,
defined in the
range of $0\leq\nu\leq 1$. 
The meaning of $\nu$ is described in details in \cite{cgs,cgst}, and we can 
briefly express it as follows.
The D5 brane as a baryon vertex
has two cusps where $U(1)$ fluxes come out the D5 form. The total 
number of the flux is $N_c$, and they
are separated to $N_c(1-\nu)$ and $N_c\nu$ to each cusp point. Namely,
the meaning of $\nu$ is then the ratio of this flux separation.

\vspace{.3cm}
\noindent{\bf Equations of motion}

Here we comment on the equations of motion which we used actually in obtaining
the baryonium state and the formal one which can be obtained from the linear
terms of the fluctuations of the fields on the brane. 
For the latter case, starting from (\ref{d5action}) or (\ref{d3action}),
the equations of motion for $r(\theta)$, $x(\theta)$ and $A_t(\theta)$ are obtained as
\bea\label{classical-1}
    \sin^4\theta\partial_r\left(e^{\Phi/2} \sqrt{\tilde{K}_{(0)}}\right)
     -\partial_{\theta}\left(\sin^4\theta {e^{\Phi/2}\over \sqrt{\tilde{K}_{(0)}}} r'\right)
         &=&0\, , \nonumber \\
    \partial_{\theta}\left(\sin^4\theta {r^4 e^{\Phi/2}\over R^4 \sqrt{\tilde{K}_{(0)}}} x'\right)
         &=&0\, ,  \nonumber \\
         \partial_{\theta}\left(\sin^4\theta {e^{-\Phi/2}\over \sqrt{\tilde{K}_{(0)}} }\tilde{A}_t'-D\right)
         &=&0\, ,
\eea 
where $r'=\partial_{\theta}r(\theta)$, $x'=\partial_{\theta}x(\theta)$ and
\beq
  \tilde{K}_{(0)}=\left(r^2+(r')^2+{r^4\over R^4} (x')^2-e^{-\Phi}\tilde{F}_{t \theta}^2\right)\, .
\eeq 
The latter two
equations are solved as
\bea\label{classical-2}
    \sin^4\theta {r^4 e^{\Phi/2}\over R^4 \sqrt{\tilde{K}_{(0)}}} x'
         &=&h\, , \nonumber \\
         \sin^4\theta {e^{-\Phi/2}\over \sqrt{\tilde{K}_{(0)}} }\tilde{A}_t'-D
         &=&0\, ,
\eea 
where $h$ is a constant and identified with the one used in our previous paper 
by the same notation \cite{GINT2}. 
Using these, we obtain the equation for $r(\theta)$, which will be solved 
numerically due to its complicated form.  

\vspace{.3cm}
In order to solve in more understandable way, it is convenient to use the
equations of motion considered before in \cite{GINT2}, and they are given in the Appendix A. It is easy to prove the equivalence of the solutions of the above
equations (\ref{classical-1}) and the one given in \cite{GINT2}.

We notice here that the parameter $h$ given in the above
equations (\ref{classical-2}) is equivalent with the one of equation (\ref{h})
in the Appendix A.
Secondly, we briefly review how our solution can be identified with the bound
state of D5 and anti-D5 branes according to the previous work.

\begin{figure}[htbp]
\begin{center}
  \includegraphics[width=9cm]{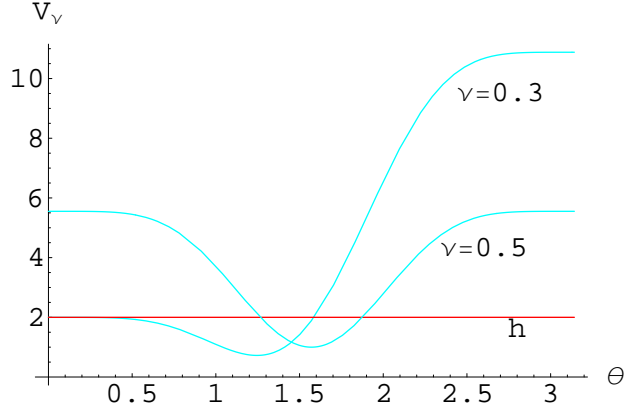}
\caption{\small $V_{\nu}(\theta)$
for $\nu=0.5$ and $0.3$. The horizontal line shows a sample line
of $V_{\nu}(\theta_0)={R^4h^2\over r_0^4+q}=2.0$. The crossing points between this line and the curves of
$V_{\nu}$ correspond to $\theta_0$ and $\theta_1$. \label{Vnu}}
\end{center}
\end{figure}


\vspace{.3cm}
\noindent{\bf Approximate solution:}

We firstly show how the region of $\theta$ in the baryonium solutions
is restriced by using $h$.
From Eq.~(\ref{h2}), we find
\beq\label{consth2}
 {p_{\theta}^2\over r^2}+p_r^2=\left({R\over r}\right)^4
\left({r^4+q\over R^4}V_{\nu}-h^2\right) \geq 0 \, ,
\eeq
then it leads to
\beq\label{consth3}
{r^4+q\over R^4}V_{\nu}\geq h^2~.
\eeq
For prototypical small energy solution, 
$r$
takes its minimum $r\simeq r_0$ at the mid point $x=0$. Expressing as 
$\theta\vert_{x=0}=\theta_0$
at this mid point, the value of $\theta_0$ is given by (\ref{consth3}) as
the solution of the following equation,
\beq\label{consth4}
{r_0^4+q\over R^4}V_{\nu}(\theta_0)=h^2~.
\eeq
This gives two solutions, $\theta_0$ and $\theta_1$ with $\theta_0>\theta_1$,
as shown in the Fig. \ref{Vnu}.
Then we find that the solutions are restricted to the region $0\leq \theta\leq \theta_1$
or $\theta_0\leq \theta\leq \pi$, and they
are identified with two different baryoniums.

\vspace{.3cm}
At the mid point, $x=0$, $\theta=\theta_0$ and $r= r_0$, which are related as
\beq\label{r0}
 r_0=\left({R^4h^2\over V_{\nu}(\theta_0)}-q\right)^{1/4}\, .
\eeq
Near this point, we can assume $\dot{r}\sim 0$
and $r\sim r_0$, and we obtain from (\ref{consth2})
\beq
 \dot{\theta}=\pm{r_0\over R^4h}\sqrt{r_0^4+q}\sqrt{V_{\nu}(\theta)-V_{\nu}(\theta_0)}
\eeq
then it is solved as 
\beq\label{baryonium}
 x(\theta)=\pm\int_{\theta_0}^{\theta}d\theta~
   {R^4h\over r_0\sqrt{r_0^4+q}\sqrt{V_{\nu}(\theta)-V_{\nu}(\theta_0)}}
\eeq
This solution is symmetric with respect to $x=0$ axis in $x$-$\theta$ plane.
The important point of this approximate
solution is that the solution runs from $x=0$ 
to two opposite directions, however they are going to the same pole on $S^5$ 
but with different values of $x$. These two 
parts are considered as D5 brane and anti-D5
brane, and they can be connected at $x=0$ by a continuous flow of the $U(1)$ flux in 
the D5 brane. 
The similar behavior has been seen in other D brane and anti-D brane bound states
\cite{KMMW2,SS,SSu}.
We regarded this as the baryonium since 
the fluxes at the two end points of this configuration have the same magnitude and 
opposite directions.

\vspace{.3cm}
\noindent{\bf Numerical solution}

\begin{figure}[htbp]
\begin{center}
  \includegraphics[width=9cm,height=6cm]{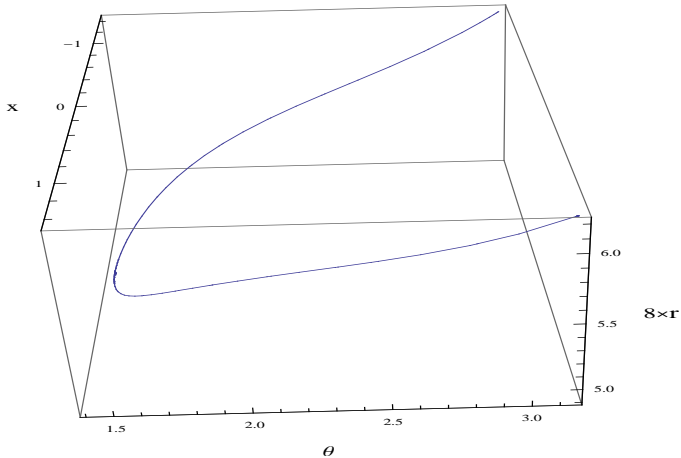}
\caption{\small A 3D representation of $D5/ \overline{D5}$ solution 
for $R=1$, $q=0.3$, $\nu=0.3$
at $h=-.67$ and $\theta_0=1.42$, where
$U$ takes its minimum and tachyon free. 
\label{sol-2}}
\end{center}
\end{figure}

In the next,
in order to see the behavior of this solution
far from $\theta_0$, we must solve the exact form of equations as
shown in \cite{GINT2}. An explicit example of the baryonium solution 
for $\nu=0.3$ is shown in the
Fig. \ref{sol-2}. This solution gives a minimum of $U$ and there is no tachyonic
mode around this configuration. So this configuration is stable as shown below. 

\vspace{.3cm}

\vspace{.3cm}
\section{Stability of the baryonium vertex}

Here we study the stability of the baryonium solution obtained above.
Previously, we have considered this problem from the energy density $U$ of the
baryonium configurations to find its minimum as a stable solution. And we have
found a minimum for $\nu=0.5$ as such an example. It has been shown  
through the contour graph in the $h-L$ plane, where
$L$ is defined as 
\beq
 L\equiv |x_+(\pi)-x_-(\pi)|\, . 
\eeq
The baryonium solutions are considered as the bound states of D5 brane and
anti-D5 brane. Then they should be separated by a definite distance to evade
tachyon modes on the D5 branes. 
The distance between these branes is characterized by $L$,
and the $U_{min}$ is found in the region of $L^2/R^4>{1\over 2\pi\alpha'}$.
This implies that the tachyonic mode of strings connecting these branes 
becomes  massive since the mass square of this
string is given as $L^2/R^4-1/(2\pi\alpha')$ \cite{Sen}.
However, we must notice near the mid point of the baryonium configuration where
the two branes are overlapping, then we might expect to find tachyons at this point.
Here, we study this problem through the fluctuation mode around 
whole region of the classical configurations.

\vspace{.3cm}
\subsection{Fluctuations}
In order to consider the fluctuations, we back to the action of D5 brane (\ref{d5action}), 
and expand it with respect to the fluctuations, $\delta r(t,\theta)=r-\bar{r},\delta x(t,\theta)=x-\bar{x}$ and $\delta A_t(\theta, t)={A}_t-\bar{A}_t$, up to their quadratic terms. Here $\bar{r}$, $\bar{A}_t(\theta)$ and $\bar{x}$
are the solutions of the equations of motion. 
{As for the $\delta A_t$, we retain it although it is not dynamical due to the lack
of the kinetic term.}
The quadratic part of (\ref{d5action}) for these fluctuations is given as
\beq\label{quadratic-3f}
 L_{(2)}=-\sin^4\theta~\left(\tilde{L}_{(2)}^{x,r}+\tilde{L}_{(2)}^{A_t}\right)\, ,
\eeq
where
\bea
  \tilde{L}_{(2)}^{x,r}&=&A_{(2)}{\delta r}^2+{B_{(0)}\over 2}\left[
  {\delta r'}^2+\left({r\over R}\right)^4{\delta x'}^2-
  \left(r^2+{r'}^2\right){\delta \dot{x}}^2 
    -\left({R^4\over r^2}+{x'}^2\right){\delta \dot{r}}^2
      +2x'r'{\delta \dot{x}}{\delta \dot{r}} \right] \nonumber \\
  && +\partial_r\left(B_{(0)}\right)r'{\delta r'}{\delta r}+
      \partial_{r}\left({r^4\over R^4}B_{(0)}\right)x'{\delta x'}{\delta r}\, \nonumber \\
  && -{B_{(0)}\over 2\tilde{G}_{(0)}}\left({r'}^2{\delta r'}^2
       +\left({r\over R}\right)^8{x'}^2{\delta x'}^2
       +2\left({r\over R}\right)^4r'x'{\delta r'}{\delta x'}\, ,
\right)
\eea
\bea
  \tilde{L}_{(2)}^{A_t}&=&-\partial_r\left(e^{-\Phi}B_{(0)}\right)A_t'{\delta A_t'}{\delta r}
      -{e^{-\Phi}B_{(0)}\over 2}{V_{\nu}\over \sin^8\theta}{\delta A_t'}^2
     +{e^{-\Phi}B_{(0)}\over \tilde{G}_{(0)}}A_t'\left(
      r'{\delta r'}+\left({r\over R}\right)^4x'{\delta x'}\right){\delta A_t'}\, \nonumber \\
 && {} 
\eea
where $r'=\partial_{\theta}r(\theta)$, $x'=\partial_{\theta}x(\theta)$,
$\dot{\delta r}=\partial_{t}\delta r$ and $\dot{x}=\partial_{t}\delta x$.
\beq
  \tilde{G}_{(0)}=\left(r^2+(r')^2+{r^4\over R^4} (x')^2\right)
{\sin^8\theta\over V_{\nu}}\, .
\eeq

\beq
  B_{(0)}=e^{\Phi/2}/\sqrt{\tilde{G}_{(0)}}\, ,\quad
  A_{(2)}={1\over 2}\partial_r^2\left(e^{\Phi/2}\sqrt{\tilde{G}_{(0)}}\right)\, ,
  \quad e^{\Phi}=1+{q\over r^4}\, .
\eeq

We notice here as mentioned above 
that $\delta A_t$ is not a dynamical fluctuation since
there is no time derivative term for it. It can be regarded as a kind of an 
auxiliary field, then we integrate it out from $L_{(2)}$. 
It is performed by rewriting $\tilde{L}_{(2)}^{A_t}$ as follows,
\beq
 \tilde{L}_{(2)}^{A_t}=-C_{(2)}\left(\delta A_t'-
          \sqrt{{\tilde{L}_{(2)}^{A_t~\rm{ind}}/ C_{(2)}}}\right)^2
           +\tilde{L}_{(2)}^{A_t~\rm{ind}}\, , \quad 
           C_{(2)}={e^{-\Phi}B_{(0)}\over 2}{V_{\nu}\over \sin^8\theta}\, ,
\eeq
where
\bea
  \tilde{L}_{(2)}^{A_t~\rm{ind}}&=&{1\over 2}{e^{2\Phi}D^2\over V_{\nu}}{\tilde{G}_{(0)}\over B_{(0)}}\left(
 -\partial_r\left(e^{-\Phi}B_{(0)}\right){\delta r}      
     +{e^{-\Phi}B_{(0)}\over \tilde{G}_{(0)}}\left(
      r'{\delta r'}+\left({r\over R}\right)^4x'{\delta x'}\right)\right)^2\, \nonumber \\
 && {} 
\eea
In order to remove $\delta A_t$, we replace $\tilde{L}_{(2)}^{A_t}$
by $\tilde{L}_{(2)}^{A_t~\rm{ind}}$ in (\ref{quadratic-3f}), then
we can restrict to two fluctuations $\delta x$ and $\delta r$ to see
the frequency of these fluctuations, which have their time derivatives.

The modified quadratic term is obtained as
\beq
\tilde{L}_{(2)}\equiv \tilde{L}_{(2)}^{x,r}+\tilde{L}_{(2)}^{A_t~\rm{ind}}
\eeq
and 
\bea
  \tilde{L}_{(2)}&=&\tilde{A}_{(2)}{\delta r}^2+{B_{(0)}\over 2}\left[
  \left({R^4\over r^2}+{x'}^2\right)\left(-{\delta \dot{r}}^2
          +\left({r\over R}\right)^4{1\over Q_{(0)}}{\delta r'}^2\right)
        +2x'r'{\delta \dot{x}}{\delta \dot{r}}\right. \nonumber \\   
 && \left.+\left(r^2+{r'}^2\right)\left(-{\delta \dot{x}}^2+\left({r\over R}\right)^4{1\over Q_{(0)}}{\delta x'}^2 \right)
      -\left({r\over R}\right)^4{1\over Q_{(0)}}2x'r'{\delta {x'}}{\delta {r'}} \right] \nonumber \\
  && +Q_{(1)}
        r'{\delta r'}{\delta r}+Q_{(2)}x'{\delta x'}{\delta r}\, 
\eea
where
\beq
 \tilde{A}_{(2)}={A}_{(2)}+{1\over 2}
   {e^{2\Phi}\tilde{G}_{(0)}\over B_{(0)}}
       {D^2\over V_{\nu}}\left(\partial_r\left(e^{-\Phi}B_{(0)}\right)\right)^2\, , \quad
   Q_{(0)}=r^2+{r'}^2+{x'}^2\left({r\over R}\right)^4    
\eeq
\beq
  Q_{(1)}=\left(\partial_r\left(B_{(0)}\right)
        -{D^2\over V_{\nu}}e^{\Phi}\partial_r\left(e^{-\Phi}B_{(0)}\right)\right)\, , \quad Q_{(2)}=\left(\partial_r\left({r^4B_{(0)}\over R^4}\right)
        -{r^4D^2\over R^4V_{\nu}}e^{\Phi}\partial_r\left(e^{-\Phi}B_{(0)}\right)\right)\, ,
\eeq
Then we can estimate the eigenvalues of the frequency of fluctuation by imposing
an appropriate boundary conditions for each eigenfunctions for these fluctuations.
These eigen functions are obtained as a solution of 
\beq\label{linear-eq}
 M_{ij}\vec{\delta f_j}=\left(\begin{array}{cc}
      M_{rr} & M_{rx}  \\
       M_{xr} & M_{xx}  
        \end{array}    
\right)
\left(\begin{array}{c}
      \delta r \\
       \delta x 
        \end{array}    
\right)=0\, \quad {\rm where} \quad 
   \vec{\delta f_j}=(\delta r, \delta x)
\eeq
where
\bea\label{quadra-ope}
  M_{rr}&=&{1\over 2}\left({R^4\over r^2}+{x'}^2\right)B_{(0)}\partial_t^2+\tilde{A}_{(2)}
         -{1\over \sin^4\theta}\partial_{\theta}\left(
   {r^4\sin^4\theta\over 2R^4}{B_{(0)}\over Q_{(0)}}\left({R^4\over r^2}+{x'}^2\right)\partial_{\theta}\right)  \nonumber \\
   &&  +Q_{(1)}r'\partial_{\theta}\, ,\\
  M_{xx}&=&{r^2+{r'}^2\over 2}B_{(0)}\partial_t^2
         -{1\over \sin^4\theta}\partial_{\theta}\left({r^4\sin^4\theta\over 2R^4}{B_{(0)}\over Q_{(0)}}\left[r^2+{r'}^2\right]\partial_{\theta}\right)\, ,\\
  M_{rx}&=&{1\over 2}\left\{
     Q_{(2)}x'\partial_{\theta}+
     {1\over \sin^4\theta}\partial_{\theta}\left({r^4\sin^4\theta\over R^4{Q}_{(0)}}B_{(0)}r'
            x'\partial_{\theta}\right)
 -r'x'B_{(0)}\partial_t^2\right\}\, , \\
  M_{xr}&=&{1\over 2}\left\{
     -{\bf {1\over \sin^4\theta}\partial_{\theta}\sin^4\theta} Q_{(2)}x'+
     {1\over \sin^4\theta}\partial_{\theta}\left({r^4\sin^4\theta\over R^4{Q}_{(0)}}B_{(0)}r'
            x'\partial_{\theta}\right)\right. \, , \nonumber \\
 &&\left. -r'x'B_{(0)}\partial_t^2\right\}\, , 
\eea
In the followings, we estimate the eigen-frequencies of the fluctuations in order
to see whether the solution is stable or not.

\vspace{.3cm}
\subsection{Estimation of eigen-frequencies}

By assuming the following form for the fluctuations,
\beq\label{derivative-co}
 \delta r(t,\theta)=e^{i\omega_r t}\phi_r(\theta),
  \quad \delta x(t,\theta)=e^{i\omega_x t}\phi_{x}(\theta)\, ,
\eeq
we estimate the value of frequency $\omega_{r,x}$ by solving the equations given in
(\ref{linear-eq}) for $\theta_0\leq\theta\leq\pi$. 
Since we are intersted in the lowest eigenvalues of $\omega_{r,x}$ to see
the stability of our solution. Such fluctuations are belonging to the
no-node wavefunctions for $\phi_{r,x}(\theta)$. Then the equations are solved
under the condition, $\phi_{r,x}(\theta_0)=$ const. and
\beq\label{Diri}
  \phi_{r,x}(\pi)=0
\eeq
or 
\beq\label{Neu}
   \partial_{\theta}\phi_{r,x}(\pi)=0
\eeq
for Diriclet or Neumann condition at the boundary $\theta=\pi$ respectively.
The constant values at $\theta_0$ are determined by the normalization condition
of the wave functions $\phi_{r,x}(\theta)$.

We calculated the eigenvalues for the classical configurations, which
provide the minimum value of $U$ for each $\nu$, since such configurations
would be stable and positive $\omega^2s$ are expected.
\begin{figure}[htbp]
\begin{center}
\hskip.05cm
\includegraphics[width=8cm,height=6cm]{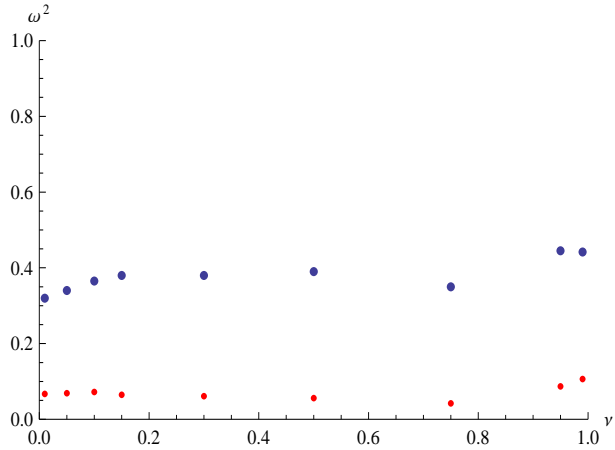}
   \caption{The lowest  for Diriclet boundary condition.
The upper (lower) points show $\omega^2_r$ ($\omega^2_x$).}
   \label{eigen-1}
   \end{center}
\end{figure}
In the Fig. \ref{eigen-1}, we show the lowest eigenvalues for the
Diriclet condition for various $\nu$. For all cases, we can see the
baryonium configurations are stable against the fluctuations.

\begin{figure}[htbp]
\begin{center}
\hskip.05cm
\includegraphics[width=7cm,height=4cm]{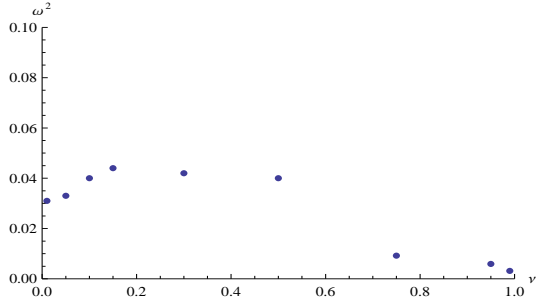}
   \caption{The lowest $\omega^2_r$ for Neumann boundary condition.}
   \label{eigen-2}
   \end{center}
\end{figure}

As for the Neumann condition, the lowest mode of $\delta x$ is the zero
mode ($\omega_x=0$) with $\phi_x(\theta)=$ const., so we show the
lowest eigenvalues of $\omega^2_r$ in the Fig. \ref{eigen-2}.
Again, we find positive $\omega^2_r$ for any $\nu$ which implies the stability
of our solutions for any $\nu$.

\vspace{.3cm}

We notice here the following points. 

(i) Firstly,
the lowest eigenvalues are obtained for the zero-node wave-
functions of each fluctuations, and the values of $\omega^2$ become large
with increasing node number of the wave-functions as expected. They are
abbreviated here.

(ii) In the second,
we show above the eigenvalues at $U=U_{min}$, however, we could find
positive eigenvalues at other values of $U$. In this sense, all the classical solutions
obtained here are stable against small fluctuations around those configurations.

(iii) The next
point to be noticed is that the above analysis is given for $0<\nu<1$, and
the point $\nu=0$ and 1 should be excluded since the configurations in these limit
of $\nu$ are not the baryonium state.

\vspace{.3cm}
\subsection{Expected configuration for $N_c=3$}

\begin{figure}[htbp]
\begin{center}
\hskip.05cm
\includegraphics[width=13cm,height=4cm]{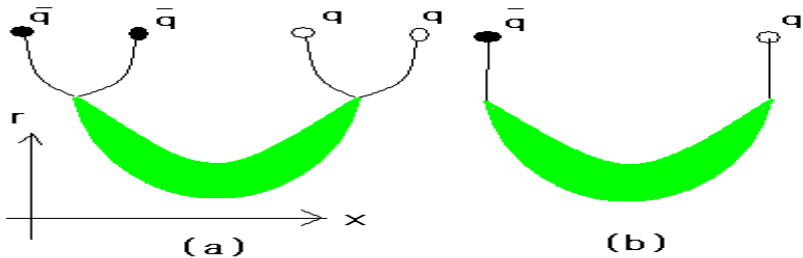}
   \caption{Expected two baryonium configurations for $N_c=3$ are shown.}
   \label{fig7}
   \end{center}
\end{figure}

The baryoniums are composed of the vertex, n quarks and n anti-quarks. 
This combination of quarks is denoted as $(n,n)$,
As a practical matter, we consider here the case
of $N_c=3$, then we find three baryonium configurations of $(3,3)$,
$(2,2)$ and $(1,1)$ quarks.
The one of $(2,2)$ and $(1,1)$ are shown in (a) and in (b)
of the Fig. \ref{fig7} respectively.
They will give a good guide for baryonium hunting in the future. 
 
\vspace{2.3cm}
\section{Stability of $D8/\overline{D8}$ and $D6/\overline{D6}$
as flavor branes}\label{eqnsec3}

In the case of type IIA string theory, 
a holographic gauge theory is considered in the D4 brane stacked
background,
\beq
ds^2_{10}=\left(\frac{U}{R}\right)^{3/2} \left(
\eta_{\mu\nu}dx^\mu dx^\nu +f(U)d\tau^2\right)+
\left(\frac{R}{U}\right)^{3/2}\left({dU^2\over f(U)}
+U^2 d\Omega_4^2\right) \ ,
\label{D4induced-metric}
\eeq
where
\beq
 f(U)=1-\left({U_{KK}\over U}\right)^3\, ,
\eeq
and $U_{KK}$ is related to the compact size of $S^1$ in the D4 world volume.
Then the flavor degrees of freedom has been introduced by embedding
$D8/\overline{D8}$ \cite{SSu} and $D6/\overline{D6}$ \cite{KMMW2}
probe branes.

\vspace{.3cm}
In these cases, the fundamental strings are not dissolved in the bound state branes.
So the fluctuations are not affected by them as in the baryonium configurations. Our
purpose to study these cases is to see the stability of those
configurations agaist the fluctuations by using our method given above. 

While, for a special case in \cite{SSu}, the stability of the configuration has been
shown as mentioned below, we can show 
their stability for more general configurations by our method. In \cite{KMMW2},
the proof of the stability of $D6/\overline{D6}$ is abbreviated, but its stability
is assured in the followings.

\vspace{.3cm}
\subsection{$D8/\overline{D8}$ branes}
For the case of $D8/\overline{D8}$ probe, they are
embedded with the following induced metric,
\beq
ds^2_{9}=\left(\frac{U}{R}\right)^{3/2} 
\eta_{\mu\nu}dx^\mu dx^\nu +\left(\left(\frac{U}{R}\right)^{3/2}f(U) 
\left({d\tau\over dU}\right)^2
+\left(\frac{R}{U}\right)^{3/2}{1\over f(U)}\right)dU^2+
  \left(\frac{R}{U}\right)^{3/2}U^2 d\Omega_4^2 \ .
\label{D8induced-metric}
\eeq
The embedded configuration of the probe brane is obtained from the D8 brane 
action,
\beq
 S_{D8}\propto \int d^4xdU U^4 (U/R)^{3/4}
   \sqrt{f(U)(\tau')^2+\left(\frac{R}{U}\right)^3{1\over f(U)}}\, ,
\eeq
where $\tau'=d\tau/dU$. The equation of motion for $\tau(U)$ is obtained as
\beq
 U^4(U/R)^{3/4}
f^{1/2}{\tau'\over \sqrt{(\tau')^2+\left(\frac{R}{U}\right)^3{1\over f^2(U)}}}=c_0\, ,
\eeq
where $c_0$ denotes an integral constant, which is taken here as
\beq
 c_0=U_0^4f^{1/2}(U_0)(U_0/R)^{3/4}\, .
\eeq
Here $U_0(\geq U_{KK})$ denotes the minimum value of $U$ for the solution. Then the solution
is given as
\beq\label{d8bard8}
 \tau(U)=\pm c_0\int_{U_0}^UdU\left(\frac{R}{U}\right)^{3/2}{1\over f(U)}
 {1\over \sqrt{U^8f(U)-c_0^2}}\, .
\eeq
As in the case of the baryonium (see Eq.(\ref{baryonium})), also this solution
can be interpreted as the bound state of D8 and anti-D8 branes which are connected
at $U=U_0$. Then we are able to study the stability of this configuration as above
by checking the existence of the tachyon on the brane. The procedure is parallel to
the case of the baryonium. By setting the fluctuation as $\delta\tau=\tau-\bar{\tau}$,
where $\bar{\tau}$ denotes the classical solution given by
Eq.(\ref{d8bard8}), the D8 action is expanded by
$\delta\tau$ as,
$$
 S_{D8}\propto \int d^4xdU U^4(U/R)^{3/4}\sqrt{f(U)G_{\tau}}\left\{
 1+{(R/U)^3\over 2G_{\tau}^2f^2(U)}(\delta\tau')^2+\right. \, 
 $$
 \beq\label{quadra-action}
 \left.  {(R/U)^3\over 2G_{\tau}f(U)}
 \eta^{\mu\nu}\partial_{\mu}\delta\tau\partial_{\nu}\delta\tau
  +\cdots \right\}\, ,
\eeq
where 
\beq\label{Gtau}
G_{\tau}=\tau'^2+(R/U)^3/f^2(U)\, .
\eeq
While, at this stage, we can see that
there is no tachyon for any $U_0$ from this form, we proceed the above procedure.
By setting as $\delta\tau=f_{\tau}(x^{\mu})\phi_{\tau}(U)$, the 
eigenmass equation of $\delta\tau$ is obtained as follows
\beq\label{d8d8eq}
 \left\{\partial_U\left(U^4\sqrt{f(U)G_{\tau}}{(R/U)^{9/4}\over G_{\tau}^2f^2(U)}
\partial_U \right)+ U^4\sqrt{f(U)G_{\tau}}{(R/U)^{9/4}\over G_{\tau}f(U)}
 m_{\tau}^2\right\}\phi_{\tau}(U)=0\, ,
\eeq
where $m_{\tau}^2$ is defined as 
\beq
\eta^{\mu\nu}\partial_{\mu}\partial_{\nu}f_{\tau}(x^{\mu})=
 m_{\tau}^2f_{\tau}(x^{\mu})\, .
\eeq
\begin{figure}[htbp]
\begin{center}
  \includegraphics[width=9cm]{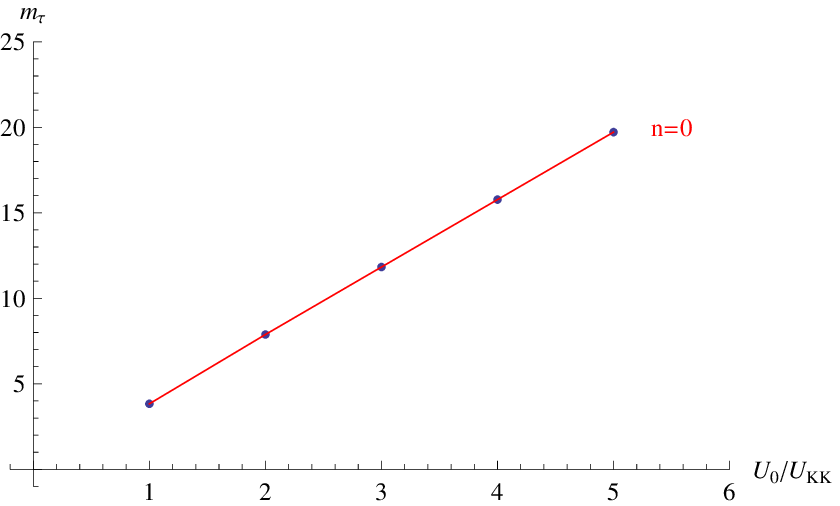}
\caption{\small The lowest value of $m_{\tau}$ versus $U_0/U_{KK}$
for $D8$-$\bar{D8}$. The values are for zero node wave-functions,
$n=0$.
\label{D8mass}}
\end{center}
\end{figure}

\vspace{.3cm}
As in the previous section, we can investigate numerically the eigen mass
$m_{\tau}^2$ of the fluctuation $\phi_{\tau}$ thorugh (\ref{d8d8eq}).
In this case, however, the boundary condition of $\phi_{\tau}$ is taken
as 
\beq
   \lim_{U\to\infty} \phi_{\tau}(U) \propto U^{-\gamma}\, , 
    \quad \gamma>{17\over 8} 
\eeq
in order to have a normalizable wave function. This is assured
from the Eq.(\ref{d8d8eq}), Under this condition, we find positive
value as the lowest mass eigenvalue $m_{\tau}^2$ for any $U_0$. 
The results are shown in the Fig. \ref{D8mass}. So we could
show the stability of this brane anti-brane bound-state also.
We should notice that $m_{\tau}$ increases linearly with $U_0/U_{KK}$
as seen from Fig.\ref{D8mass}.

\vspace{.3cm}
Finally we notice that the stability of this configuration has been studied
for the case $U_0=U_{KK}$ \cite{SSu}. 
However, here, we could show its stability for any
value of $U_0$, $U_0>U_{KK}$.
{The reason why this bound state is stable for any $U_0$, where D8 and anti-D8
branes are connected, is that there is no $U(1)$ flux in the brane in this case.
This phenomenon seems to be universal for any bound state
of D and anti-D branes.}

\vspace{1.3cm}
\subsection{$D4/D6$-$\bar{D6}$ branes}


Instead of (\ref{D4induced-metric}), the D4 background solution 
is given in \cite{KMMW2} as follows 
\beq
ds^2_{10}=\left(\frac{U}{R}\right)^{3/2} 
\left(\eta_{\mu\nu}dx^\mu dx^\nu +f(U) d\tau^2\right)
+K(\rho)\left(d\lambda^2+\lambda^2 d\Omega_2+dr^2+r^2 d\phi^2\right) \ .
\label{D4background}
\eeq
\beq
U(\rho)=\left(\rho^{3/2}+{U_{KK}^{3}\over 4\rho^{3/2}}\right)^{2/3}\, ,
\eeq
where $\rho^2=\lambda^2+r^2$ and
\beq
e^{\Phi}=g_s \left(\frac{U}{R}\right)^{3/4}\, , \quad 
f(U)=1-\left({U_{KK}\over U}\right)^3\, , \quad 
K(\rho)={R^{3/2}U^{1/2}\over \rho^2}\, .
\eeq

The D6 is embedded in the following metric,
\beq
ds^2_{7}= 
\left(\frac{U}{R}\right)^{3/2}\eta_{\mu\nu}dx^\mu dx^\nu +
R^{3/2}U^{1/2} d\Omega_2^2 +
\left[\left(\frac{U}{R}\right)^{3/2}f(U){\tau'}^2+{1\over\left(\frac{U}{R}\right)^{3/2}f(U)}\right] dU^2\ .
\label{type2A-1}
\eeq
Here we supposed that the profile is determined by $\tau(U)$. 
Then the DBI action of D6 brane is given by
\beq\label{d6}
 L_{D6}=T_{D6}R^{3/2}~\left({U\over R}\right)^{17/4}\sqrt{f(U)G_{\tau}}\, .
\eeq
where $G_{\tau}$ is given in Eq.(\ref{Gtau}) above. Solving the equation
of motion derived from $L_{D6}$,
we obtain $D4/D6$-$\bar{D6}$ embedded solution $\tau(U)$
in the background (\ref{D4background}) as in the $D8$ case.
The equation and its solution are given as follows,
\beq
 (U/R)^{17/4}
f^{1/2}{\tau'\over \sqrt{G_{\tau}}}=d_0\, ,
\eeq
where the integral constant $d_0$ is given as
\beq
 d_0=f^{1/2}(U_0)(U_0/R)^{17/4}\, .
\eeq
Here $U_0$ is the minimum value of $U$ and $\tau'(U_0)=\infty$, and
the solution
is given as
\beq\label{d6bard6}
 \tau(U)=\pm \int_{U_0}^UdU\left(\frac{R}{U}\right)^{3/2}{1\over f(U)}
 {1\over \sqrt{(U/R)^{17/4}f(U)/d_0^2-1}}\, .
\eeq
\begin{figure}[htbp]
\begin{center}
  \includegraphics[width=9cm]{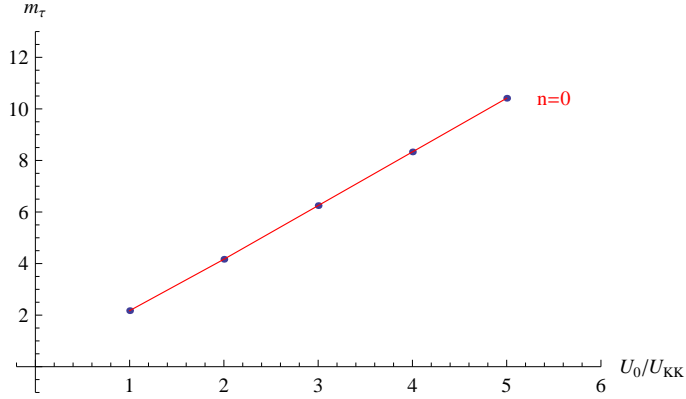}
\caption{\small The lowest value of $m_{\tau}$ versus $U_0/U_{KK}$
for $D6$-$\bar{D6}$. The values are for zero node wave-functions,
$n=0$.
\label{D6mass}}
\end{center}
\end{figure}

\vspace{.3cm}
As for the fluctuation of $\tau$, we can see its quadratic term by expanding
(\ref{d6}) with respect to $\delta\tau(=\tau-\tau_{\rm cl})$ as follows,
$$
 S_{D6}\propto \int d^4xdU (U/R)^{17/4}\sqrt{f(U)G_{\tau}}\left\{
 1+{(R/U)^3\over 2G_{\tau}^2f^2(U)}(\delta\tau')^2+\right. \, 
 $$
 \beq\label{quadra-action}
 \left.  {(R/U)^3\over 2G_{\tau}f(U)}
 \eta^{\mu\nu}\partial_{\mu}\delta\tau\partial_{\nu}\delta\tau
  +\cdots \right\}\, ,
\eeq
By setting as $\delta\tau=f_{\tau}(x^{\mu})\varphi_{\tau}(U)$, the 
eigenmass equation of $\varphi_{\tau}(U)$ is obtained as follows
\beq\label{d6d6eq}
 \left\{\partial_U\left(\sqrt{f(U)G_{\tau}}{(U/R)^{5/4}\over G_{\tau}^2f^2(U)}
\partial_U \right)+ \sqrt{f(U)G_{\tau}}{(U/R)^{5/4}\over G_{\tau}f(U)}
 m_{\tau}^2\right\}\varphi_{\tau}(U)=0\, ,
\eeq
where the mass eigen-value $m_{\tau}^2$ is defined as above,
$
\eta^{\mu\nu}\partial_{\mu}\partial_{\nu}f_{\tau}(x^{\mu})=
 m_{\tau}^2f_{\tau}(x^{\mu})\,$ .
The linear term of $\delta\tau$ disappears due to the field equation
of $\tau$.

In solving (\ref{d6d6eq}), we must impose the boundary condition for
$\varphi$ from the normalizability as
\beq
   \lim_{U\to\infty} \varphi_{\tau}(U) \propto U^{-\gamma_1}\, , 
    \quad \gamma_1>{7\over 8}\, .
\eeq
Under this condition, we have find positive mass eigen-value $m_{\tau}^2$
as in the case of $D8$ branes. 
The results are shown in the Fig.\ref{D6mass}. In this case also,
$m_{\tau}$ increases linearly with $U_0/U_{KK}$
as seen in the case of $D8$ branes. In any case,
the $D6/\bar{D6}$ state is also stable
against the fluctuation of $\tau(U)$.

\section{Summary}\label{comparison}

We have examined the stability of D brane anti-D brane bound state 
embedded as a probe in some 10D supergravity background. We have restricted 
our attention to some special configurations, the baryonium state in type
IIB theory and flavor branes in type IIA theory.

The vertex of the baryonium has been obtained as a bound state of D5 and anti-D5 brane in type IIB theory 
in terms of the D5 brane action,
which wrapped on $S^5$ in the 10D AdS$_5\times S^5$ and 
gives baryon vertex at the same time. The configuration of the baryonium is
determined by the main two parameters, the gauge condensate $q$, which would
determine the enrgy scale, and the flux number
$(1-\nu)N_c$ (or $\nu N_c$), which is proportional to the quark number in the
baryonium. 

Previously, we could find an appropriate solutions of the baryonium which
gives a minimum of the classical action for any value of $\nu$ and $q$. In 
order to confirm the stability of this configuration furthermore,
this time, we have studied the time-frequency $\omega$ of the fluctuations
for the relevant fields living 
on the D5 brane to search for the unexpected tachyonic modes. 
{The eigen value $\omega$ has been investigated numerically 
for the eigen functions $\phi_r(\theta)$ and $\phi_x(\theta)$ on which 
the Dirichlet or Neumann boundary conditions are imposed. For both
bondary conditions and at any value of $\nu$,
we have found non-negative $\omega^2$ for the configurations given 
at $U=U_{min}$. Surprisingly, the non-negative $\omega^2$ is observed even if
we consider the fluctuations around 
any configurations of $U(>U_{min})$. In this sense, any baryonium
configuration given as the classical solution would be stable agaist the
fluctuations around the baryonium solutions considered here. 


As in the case of the baryon, we must add quarks to the vertex to find a
complete baryonium state. The method would be parallel to the case of the baryon
\cite{GI}, where the no force conditions are needed at each cusp to connect
$(1-\nu)N_c$ fundamental strings. this will be done in the near future.

We have also investigated the stability of the bound-state of $D8/\overline{D8}$ 
and $D6/\overline{D6}$, which are introduced as the brane for the fundamental quarks
in the type IIA theory. We found that these bound states are stable. This stability
would be promised by the fact that there are no $U(1)$ flux dissolved in the branes
in these cases.

\section*{Acknowledgements}
The authors would like to thank M. Ishihara and T. Taminato for useful discussions.

\section*{Appendix}
\section*{A: Comment on Equation of motion}
And the action is rewritten by eliminating the gauge field
in terms of the solution for $\tilde{F}_{t \theta}$ given by the second equation
of (\ref{classical-2}) to obtain an energy
functional of the embedding coordinate only 
\footnote{
$U$ is obtained by a Legendre transformation of $L$, which is defined
as $S=\int dt L$, as $U={\partial L\over \partial\tilde{F}_{t \theta}}
\tilde{F}_{t \theta}-L$. Then equations of motion of (\ref{d3action}) 
provides the same solutions of the one of $U$. 
}:
\be \label{u}
U = {N\over 3\pi^2\alpha'}\int d\theta~e^{\Phi/2}
\sqrt{r^2+r^{\prime 2} +(r/R)^{4}x^{\prime 2}}\,
\sqrt{V_{\nu}(\theta)}~.
\ee
\be\label{PotentialV}
V_{\nu}(\theta)=D(\nu,\theta)^2+\sin^8\theta
\ee
where we used $T_5 \Omega_{4}R^4=N/(3\pi^2\alpha')$.

\vspace{.3cm}
We notice that the solution for $x(\theta)$ in our baryonium configuration is given by a
two valued functions for the variable $\theta$, which is restricted to
(i) $\theta_0\leq \theta\leq \pi$ or (ii) $0\leq \theta\leq \theta_1$
as explained below. So it is convenient to solve the equations by changing
the variable from $\theta$ to $x$ in $U$ as,
\be \label{upar}
U = {N\over 3\pi^2\alpha'}\int dx~e^{\Phi/2}
\sqrt{ r^2\dot{\theta}^2+\dot{r}^2+(r/R)^{4}}~
\sqrt{V_{\nu}(\theta)},
\ee
where dots denotes the derivative with respect to $x$. 
In this form, $x$ is not contained explicitly, then
we can introduce an integral constant $h$ as
a ``Hamiltonian'' for the corresponding ``time variable'' $x$ as follows
\beq \label{h}
 h=\dot{r}p_{r}+\dot{\theta}p_{\theta}-L\, ,\quad
\eeq
where 
$ L=e^{\Phi/2} \sqrt{ r^2\dot{\theta}^2+\dot{r}^2+(r/R)^{4}}~\sqrt{V_{\nu}(\theta)}
$ , 
\beq
 p_{r}={\partial L\over \partial \dot{r}}=\dot{r}Q\, , \quad
 p_{\theta}={\partial L\over \partial \dot{\theta}}=r^2\dot{\theta}
           Q\, , \quad Q=\left({R\over r}\right)^2\sqrt{e^{\Phi}V_{\nu}-
   \left({p_{\theta}^2\over r^2}+p_{r}^2\right)}\, .
\eeq
Then $h$ is rewritten in terms of the momentum as
\beq\label{h2}
 h 
=-\left({r\over R}\right)^2\sqrt{e^{\Phi}V_{\nu}-
   \left({p_{\theta}^2\over r^2}+p_{r}^2\right)}\, ,
\eeq
and the Hamilton equations of motion are obtained as
\beq\label{dot-r-th}
\dot{r}=
{p_r\over Q}\, , \quad
\dot{\theta}=
{p_{\theta}\over r^2Q}\, , \quad
\dot{p}_r=-{\partial h\over \partial r}\, , \quad
 \dot{p}_{\theta}=-{\partial h\over \partial \theta}~.
\eeq
These equations are convenient to find the baryonium vertex solutions.

\vspace{.5cm}

\end{document}